# Another derivation of Weinberg's formula

**Zhi-Yong Wang**, **Cai-Dong Xiong**

*School of Optoelectronic Information, University of Electronic Science and Technology of China, Chengdu 610054, CHINA*

**Abstract**

To investigate how quantum effects might modify special relativity, we will study a Lorentz transformation between classical and quantum reference frames and express it in terms of the four-dimensional (4D) momentum of the quantum reference frame. The transition from the classical expression of the Lorentz transformation to a quantum-mechanical one requires us to symmetrize the expression and replace all its dynamical variables with the corresponding operators, from which we can obtain the same conclusion as that from quantum field theory (given by Weinberg's formula): owing to the Heisenberg's uncertainty relation, a particle (as a quantum reference frame) can propagate over a spacelike interval.

## 1. Introduction

Special relativity has been developed on the basis of classical mechanics without taking into account any quantum-mechanical effect, which implies that some traditional conclusions in special relativity might be modified on condition that quantum-mechanical effects cannot be ignored. For example, special relativity tells us that any particle cannot propagate over a spacelike interval, but according to quantum field theory, such superluminal behavior does actually exist [1-4]. In particular, Steven Weinberg has presented a detailed explanation for this superluminal propagation [5]. In Ref. [5] Weinberg discussed as follows (with some different notations and conventions):

Although the relativity of temporal order raises no problems for classical physics, it plays a profound role in quantum theories. The uncertainty principle tells us that when we specify that a particle is at position $x_1$ at time $t_1$, we cannot also define its velocity precisely. In consequence there is a certain chance of a particle getting from $(t_1, x_1)$ to $(t_2, x_2)$ even if the spacetime interval is spacelike, that is, $|x_1 - x_2| > c|t_1 - t_2|$. To be more precise, the probability of a particle reaching $(t_2, x_2)$ if it starts at $(t_1, x_1)$ is nonnegligible as long as (we call Eq. (1) *Weinberg's formula*)

$$0 < (x_1 - x_2)^2 - c^2(t_1 - t_2)^2 \leq (\hbar/mc)^2, \tag{1}$$

where $\hbar$ is Planck's constant (divided by $2\pi$), c is the velocity of light in vacuum, and $m$ is the particle's mass (and then $\hbar/mc$ is the Compton wavelength of the particle). For simplicity, let $t_1 = 0$, $x_1 = (0,0,0)$, $t_2 = t$, and $x_2 = (x,0,0)$, for the moment Weinberg's formula (1) can be rewritten as ($\lambda \equiv \hbar/mc$ denotes the Compton wavelength of the particle)

$$0 > c^2 t^2 - x^2 \geq -(\hbar/mc)^2 = -\lambda^2. \tag{2}$$

We are thus faced again with our paradox; if one observer sees a particle emitted at $(t_1, x_1) = (0,0,0,0)$, and absorbed at $(t_2, x_2) = (t, x, 0, 0)$, and if $c^2 t^2 - x^2$ is negative (but greater than or equal to $-(\hbar/mc)^2$), then a second observer may see the particle absorbed at $x_2 = (x,0,0)$ at a time $t_2 = t$ before the time $t_1 = 0$ it is emitted at $x_1 = (0,0,0)$. There is only one known way out of this paradox. The second observer must see a particle emitted at $x_2 = (x,0,0)$ and absorbed at $x_1 = (0,0,0)$. But in general the particle seen by the second observer will then necessarily be different from that seen by the first observer (it is the antiparticle of the particle seen by the first observer). In other words, to avoid a possible causality paradox, one can resort to the particle-antiparticle symmetry. The process of a particle created at $(t_1, \boldsymbol{x}_1)$ and annihilated at $(t_2, \boldsymbol{x}_2)$ as observed in a frame of reference, is identical with that of an antiparticle created at $(t_2, \boldsymbol{x}_2)$ and annihilated at $(t_1, \boldsymbol{x}_1)$ as observed in another frame of reference.

In fact, Weinberg's above argument is equivalent to the usual argument in quantum-field-theory textbooks: let the probability amplitude for a particle propagating over a spacelike interval $(x-y)^2 < 0$ be denoted as $D(x-y) = \langle 0 | \phi(x)\phi(y) | 0 \rangle$, correspondingly, the quantity $D(y-x) = \langle 0 | \phi(y)\phi(x) | 0 \rangle$ represents the probability amplitude for the corresponding antiparticle propagating backwards over the spacelike interval. Because $D(x-y) = D(y-x)$ for $(x-y)^2 < 0$, the two spacelike processes are undistinguishable and the commutator $[\phi(x), \phi(y)] = \langle 0 | [\phi(x), \phi(y)] | 0 \rangle = 0$, such that the causality is maintained.

Weinberg's formula given by Eq. (1) or (2) comes from a rough estimate. In Section 3, a more rigorous form of Weinberg's formula will be obtained within the framework of quantum field theory. In this letter, we will investigate how quantum effects might modify special relativity by studying quantum-mechanical Lorentz transformation, from which we will obtain Weinberg's formula at the first-quantized level. Our conclusion is also valid for photon tunneling, because guided photons inside a waveguide can be treated as massive particles.

**2. Quantum-mechanical Lorentz transformation**

One can combine special relativity with quantum mechanics via two different approaches: (1) developing quantum mechanics on the basis of special relativity, one can obtain relativistic quantum theory (including relativistic quantum mechanics and quantum field theory); (2) developing special relativity on the basis of quantum mechanics, one might obtain a quantum-mechanical special relativity. The former has been successful, while the latter remains to be achieved. Historically, many attempts have been made to investigate how quantum effects might modify special relativity (e.g., try to apply quantum-mechanical uncertainty to the reference frames of relativity; try to extend the concept of macroscopic observers to include that of quantum observers, etc.) [6-8], a quantum reference frame defined by a material object subject to the laws of quantum mechanics has been studied [9-13]. However, these attempts have not been completely successful. For example, in Ref. [9] quantum reference frame has been discussed within the framework of nonrelativistic quantum theory, such that it has been concerned with Galilean relativity, instead of Einstein relativity. Furthermore, to take a "quantum special relativity" as being a limit of quantum gravity in a similar way Special Relativity is a limit of General Relativity, Doubly Special Relativity has been proposed [14-18], whose idea is that there exist in nature two observer-independent scales, of velocity, identified with the speed of light, and of mass, which is expected to be of order of Planck mass. However, even if Doubly Special Relativity is valid, it does not deviate from the usual Special Relativity unless the scale under consideration approaches the Planck scale, and thus it has nothing to do with our present issue.

Consider that Lorentz transformations are the base of special relativity, to investigate how quantum effects might modify special relativity, we will study a Lorentz transformation between classical and quantum reference frames and express it in terms of the four-dimensional (4D) momentum of the quantum reference frame.

Consider two inertial reference frames $S$ and $S'$ with a relative velocity $\mathbf{v} = (v, 0, 0)$ between them. We shall denote observables by unprimed variables when referring to $S$, and by primed variables when referring to $S'$, and then the time and space coordinates of a point are denoted as $(t, x, y, z)$ and $(t', x', y', z')$ in the frames $S$ and $S'$, respectively. The coordinate axes in the two frames are parallel and oriented so that the frame $S'$ is moving in the positive x direction with speed $v > 0$, as viewed from $S$. Let the origins of the coordinates in $S$ and $S'$ be coincident at time $t = t' = 0$. All statements here are presented from the point of view of classical mechanics, or, in other words, they are valid in the sense of quantum-mechanical average.

From the physical point of view, a frame of reference is defined by a material object of the same nature as the objects that form the system under investigation and the measuring instruments [10]. If the mass of the material object is finite, the corresponding reference frame (say, quantum reference frame) would be subject to the laws of quantum mechanics, and the interaction between object and measuring device might not be neglected. In particular, Heisenberg's uncertainty relations forbid the exact determination of the relative position and velocity of quantum reference frame. For simplicity, we assume that the interaction between a physical system and measuring device is so small that all quantum reference frames can approximatively be regarded as inertial ones (they are inertial ones in the sense of quantum-mechanical average).

To study whether a particle can propagate over a spacelike interval, we assume that the frame $S'$ is attached to a particle Q with rest mass *m* (i.e., a quantum-mechanical object of finite mass), such that the frame $S'$ can be regarded as consisting of a measuring device and the particle Q. For simplicity, we assume that the mass of the measuring device can be ignored as compared with that of the particle Q. As a result, the frame $S'$ can approximatively be defined by the particle Q with rest mass $m$, wherein a Cartesian coordinate system is chosen in such a manner that the coordinates of the particle Q is $(t', x', 0, 0)$ as viewed in $S'$, and is $(t, x, 0, 0)$ as viewed in the frame $S$.

On the other hand, for convenience we assume that the frame $S$ has an infinite mass. In other words, the frame $S$ is a classical reference frame while the frame $S'$ is a quantum one. For simplicity, from now on we will omit the y- and z- axes. According to the Lorentz transformation one has

$$\begin{cases} x' = (x - vt) / \sqrt{1 - (v^2/c^2)} \\ t' = [t - (vx/c^2)] / \sqrt{1 - (v^2/c^2)} \end{cases}, \quad (3)$$

Because the frame $S'$ is attached to the particle Q, let $\boldsymbol{p} = (p, 0, 0)$ and $E$ denote the momentum and energy of the particle Q as observed in the frame $S$, respectively, then $p = Ev/c^2 > 0$. In other words, as observed in $S$, the particle Q has the 4D momentum $(E, p, 0, 0)$ and the 4D coordinate $(t, x, 0, 0)$. Using $E^2 = p^2c^2 + m^2c^4$ and $v = pc^2/E$, Eq. (3) can be rewritten as

$$\begin{cases} x' = (Ex - c^2 pt) / mc^2 \\ t' = (Et - px) / mc^2 \end{cases}. \quad (4)$$

As we know, the transition from the classical expression (4) to a quantum-mechanical one requires us to symmetrize Eq. (4) and replace all its variables with the corresponding operators, in such a way we formally give a quantum Lorentz transformation (in the position-space representation)

$$\begin{cases} x' = [(\hat{H}x + x\hat{H}) - c^2(\hat{p}t + t\hat{p})]/2mc^2 \\ t' = [(\hat{H}t + t\hat{H}) - (\hat{p}x + x\hat{p})]/2mc^2 \end{cases}. \tag{5}$$

where $\hat{H}$ is the Hamilton operator satisfying $\hat{H}^2 = \hat{p}^2c^2 + m^2c^4$ and $\hat{p} = -i\hbar\partial/\partial x$. Using $\mathrm{d}t/\mathrm{d}t = \partial t/\partial t + (i/\hbar)[\hat{H},t] = \partial t/\partial t = 1$ one has

$$[\hat{H},t] = \hat{H}t - t\hat{H} = 0. \tag{6}$$

That is, in contrast with the conjugate pair $x$ and $\hat{p} = -i\hbar\partial/\partial x$, $\hat{H}$ and $t$ do not constitute a conjugate pair. Likewise, owing to $\partial t/\partial x = 0$, one has $t\hat{p} = \hat{p}t$. Therefore, as viewed in the classical reference frame $S$, time coordinate $t$ acts as a parameter rather than an operator, which is in agreement with the traditional conclusion (as a result, time in quantum mechanics has been a controversial issue since the advent of quantum theory). Using $\hat{H}t = t\hat{H}$ and $t\hat{p} = \hat{p}t$ the quantum Lorentz transformation (5) can be rewritten as

$$\begin{cases} x' = \dfrac{(\hat{H}x + x\hat{H})}{2mc^2} - \dfrac{t\hat{p}}{m} \\ t' = \dfrac{t\hat{H}}{mc^2} - \dfrac{(\hat{p}x + x\hat{p})}{2mc^2} \end{cases}. \tag{7}$$

Consider that the particle Q moves relative to the frame $S$ with constant velocity $v$ along x-axis, one has $\mathrm{d}v/\mathrm{d}t = \mathrm{d}^2x/\mathrm{d}t^2 = 0 = (i/\hbar)[\hat{H},(i/\hbar)[\hat{H},x]]$, i.e., $\hat{H}[\hat{H},x] = [\hat{H},x]\hat{H}$, it follows that

$$[\hat{H}^2,x] = \hat{H}[\hat{H},x] + [\hat{H},x]\hat{H} = 2\hat{H}[\hat{H},x]. \tag{8}$$

. On the other hand, using $\hat{H}^2 = \hat{p}^2c^2 + m^2c^4$ one has

$$[\hat{H}^2,x] = \hat{p}[\hat{p},x]c^2 + [\hat{p},x]\hat{p}c^2 = -2i\hbar\hat{p}c^2. \tag{9}$$

Combining Eq. (8) with Eq. (9), one has:

$$[\hat{H},x] = -i\hbar\hat{H}^{-1}\hat{p}c^2. \tag{10}$$

From Eq. (10) one can obtain the desired result $\mathrm{d}x/\mathrm{d}t = (i/\hbar)[\hat{H},x] = \hat{H}^{-1}\hat{p}c^2$, which is related to the classical expression $v = dx/dt = pc^2/E$ and in agreement with Ehrenfest's theorems. In fact, take Dirac electron for example, by splitting up every operator into an even and an odd part so as to throw off the zitterbewegung part [19], one can obtain a *true* velocity operator that is similar to $dx/dt = \hat{H}^{-1}\hat{p}c^2$.

**3. Spacelike propagation on account of quantum-mechanical effects**

Applying $\hat{H}t = t\hat{H}$, $t\hat{p} = \hat{p}t$, $\hat{p}x = x\hat{p} - i\hbar$, $\hat{H}x = x\hat{H} - i\hbar\hat{H}^{-1}\hat{p}c^2$, $\hat{H}^2 = \hat{p}^2c^2 + m^2c^4$, $\hat{H}\hat{p} = \hat{p}\hat{H}$, $xt = tx$, and Eq. (10), one can obtain (see **Appendix A**):

$$c^2t'^2 - x'^2 = c^2t^2 - x^2 + \hbar^2c^2\hat{H}^{-2}/4. \qquad (11)$$

Owing to $\hat{H}^2 = \hat{p}^2c^2 + m^2c^4 \geq m^2c^4$ (in the sense of eigenvalues or quantum-mechanical averages of operators), for a timelike or lightlike interval $c^2t'^2 - x'^2 \geq 0$, using Eq. (11) one has

$$c^2t^2 - x^2 \geq -\hbar^2c^2\hat{H}^{-2}/4 \geq -\hbar^2/4m^2c^2 = -(\lambdabar/2)^2, \qquad (12)$$

where $\lambdabar = \hbar/mc$ is the Compton wavelength of the particle Q. Eq. (12) implies that, as observed in $S$, the particle Q can propagate over a spacelike interval provided that

$$0 > c^2t^2 - x^2 \geq -(\hbar/2mc)^2 = -(\lambdabar/2)^2, \qquad (13)$$

which is in agreement with Weinberg's formula given by Eq. (2) but for a factor of 1/4. In the following we will show that, actually, Eq. (13) is the more rigorous form of Weinberg's formula, while Eq. (2) comes from a rough estimate. To do so, let us derive Weinberg's formula within the framework of quantum field theory as follows: For simplicity let $\varphi(x) = \varphi(t, \boldsymbol{x})$ represent a scalar field operator and $|0\rangle$ denote the field's vacuum state. According to quantum field theory, the quantity

$$\Gamma(t_2 - t_1, \boldsymbol{x}_2 - \boldsymbol{x}_1) \equiv \langle 0|\varphi(t_2, \boldsymbol{x}_2)\varphi(t_1, \boldsymbol{x}_1)|0\rangle, \qquad (14)$$

represents a transition probability amplitude from the quantum state $\varphi(t_1, \boldsymbol{x}_1)|0\rangle$ to the quantum state $\varphi(t_2, \boldsymbol{x}_2)|0\rangle$, or equivalently, it represents a probability amplitude for a scalar particle to propagate from $(t_1, \boldsymbol{x}_1)$ to $(t_2, \boldsymbol{x}_2)$ [3, 4], such that the quantity $|\Gamma(t_2 - t_1, \boldsymbol{x}_2 - \boldsymbol{x}_1)|^2$ represents the corresponding probability (density). For simplicity, let $t_1 = 0$, $\boldsymbol{x}_1 = (0,0,0)$, $t_2 = t$, and $\boldsymbol{x}_2 = (x,0,0)$, and denote $\Gamma(t_2 - t_1, \boldsymbol{x}_2 - \boldsymbol{x}_1) = \Gamma(t, x)$. For our purpose, let the scalar particle be the particle Q discussed before (it happened that Weinberg took the scalar $\pi$ meson for example in Ref. [5]), then $\Gamma(t, x)$ denotes the probability amplitude for the particle Q to propagate from $(0,0)$ to $(t, x)$. According to quantum field theory, in our case one has (up to a constant factor)

$$\Gamma(t,x) = \int_{-\infty}^{+\infty} \frac{\mathrm{d}p}{2\pi} \frac{c}{2E_p} \exp[-\mathrm{i}(E_p t - px)/\hbar], \tag{15}$$

where $E_p = \sqrt{p^2 c^2 + m^2 c^4}$. Let $H_0^{(2)}(z)$ denote the zero-order Hankel function of the second kind, as the spacetime interval is spacelike (i.e., $c^2 t^2 - x^2 < 0$), one can prove that,

$$\Gamma(t,x) = (-\mathrm{i}/4) H_0^{(2)}(-\mathrm{i}\sqrt{x^2 - c^2 t^2}/\lambdabar), \tag{16}$$

where $\lambdabar = \hbar/mc$ is the Compton wavelength of the particle Q. Therefore, the asymptotic behaviors of $\Gamma(t, x)$ are governed by the Hankel function of imaginary argument, that is: $\Gamma(t, x)$ falls off like $\sqrt{1/z}\exp(-z)$ for $z = \sqrt{x^2 - c^2 t^2}/\lambdabar \to +\infty$, while falls off faster than $\sqrt{1/z}\exp(-z)$ for the other $z = \sqrt{x^2 - c^2 t^2}/\lambdabar$. As a result, $\Gamma(t, x)$ is always ignored for $z = \sqrt{x^2 - c^2 t^2}/\lambdabar > 1$, that is, one always takes $\Gamma(t, x)$ as,

$$\Gamma(t,x) \begin{cases} = 0, & \text{for } c^2 t^2 - x^2 < -(\hbar/mc)^2 = -\lambdabar^2 \\ \neq 0, & \text{for } 0 > c^2 t^2 - x^2 \geq -(\hbar/mc)^2 = -\lambdabar^2 \end{cases}. \tag{17}$$

Therefore, for the spacelike interval $c^2 t^2 - x^2 < 0$, the probability amplitude $\Gamma(t, x)$ for the particle Q to propagate from $(0,0)$ to $(t, x)$ is nonnegligible as long as Weinberg's formula given by Eq. (2) is satisfied. Then we obtain Eq. (2).

However, in addition to Eq. (2), within the framework of quantum field theory Weinberg's formula can be reformulated in a more rigorous way. In fact, in the observable sense, one should concern the probability (density) $|\Gamma(t,x)|^2$ rather than the probability amplitude $\Gamma(t,x)$, because the latter is not an observable quantity. Likewise, Eq. (16) implies that $|\Gamma(t,x)|^2$ falls off like $\exp(-2z)/z$ for $z = \sqrt{x^2 - c^2t^2}/\lambdabar \to +\infty$, while falls off faster than $\exp(-2z)/z$ for the other $z = \sqrt{x^2 - c^2t^2}/\lambdabar$. As a result, one conventionally ignores $|\Gamma(t,x)|^2$ for $2z = 2\sqrt{x^2 - c^2t^2}/\lambdabar > 1$, which gives us a criterion to estimate when the probability $|\Gamma(t,x)|^2$ can be ignored and when it cannot. For the moment, one has

$$|\Gamma(t,x)|^2 \begin{cases} = 0, \text{ for } c^2t^2 - x^2 < -(\hbar/2mc)^2 = -(\lambdabar/2)^2 \\ \neq 0, \text{ for } 0 > c^2t^2 - x^2 \geq -(\hbar/2mc)^2 = -(\lambdabar/2)^2 \end{cases}. \quad (18)$$

That is, even if $c^2t^2 - x^2$ is spacelike, the probability $|\Gamma(t,x)|^2$ for the particle Q to propagate from $(0,0)$ to $(t,x)$ cannot be ignored as long as the formula of $0 > c^2t^2 - x^2 \geq -(\lambdabar/2)^2$ is satisfied. Then a more rigorous expression of Weinberg's formula, i.e., $0 > c^2t^2 - x^2 \geq -(\lambdabar/2)^2$, is obtained within the framework of quantum field theory, it is exactly Eq. (13).

Therefore, Weinberg's formula obtained within the framework of quantum field theory (i.e., at the second-quantized level), can be also obtained via quantum-mechanical Lorentz transformation (i.e., at the first-quantized level). Where, within the framework of quantum mechanics, the rigorous result is Eq. (13); within the framework of quantum field theory, the rigorous result is Eq. (16), while Eq. (13) is obtained via Eq. (18), which corresponds to an approximation of Eq. (16). This can be due to the fact that, quantum mechanics is an approximation of quantum field theory.

Though a particle can propagate over a spacelike interval, Einstein's causality is preserved: a commutator between two observables for a spacelike interval must vanish, such that a measurement performed at one point cannot affect another measurement at a point separated from the first with a spacelike interval. Moreover, via Refs. [1-3] one can show that the particle Q satisfying Eq. (13) can correspond to the one tunneling through a potential barrier (including photons tunneling through an undersized waveguide, note that guided photons inside a waveguide can be treated as massive particles).

**4. Conclusions and discussions**

As a purely quantum-mechanical effect, the presence of the term $\hbar^2 c^2 \hat{H}^{-2}/4$ in Eq. (11) is essentially due to the commutation relation $[x, \hat{p}] = i\hbar$. Therefore, the fact that a particle with finite mass can propagate over a spacelike interval attributes to the Heisenberg's uncertainty relation. By analyzing how quantum effects might modify special relativity, we obtain Weinberg's formula at the first-quantized level.

Note that in our case, spacetime coordinates are also spacetime intervals (with respect to origins of coordinates). As mentioned before, as viewed in the classical reference frame $S$ one has $xt = tx$, and time enters as a parameter rather than an operator. On the other hand, one can prove that:

$$x't' - t'x' = -i\hbar(\hat{H}^{-1}x + x\hat{H}^{-1})/2. \qquad (19)$$

That is, as viewed in the quantum reference frame $S'$, the spacetime coordinates of the particle Q are noncommutative and time enters as an operator. In fact, once time enters as an operator, spacetime coordinates may become noncommutative. For example, let $p = mu$, by quantizing the classical expression $t = \pm x/u = \pm mx/p$ one can obtain the nonrelativistic free-motion time-of-arrival operator $\hat{T}_{non} = \pm m(\hat{p}^{-1}x + x\hat{p}^{-1})/2$ [20-24]. If we take $\hat{T}_{non} = m(\hat{p}^{-1}x + x\hat{p}^{-1})/2$, and note that in the momentum space representation one has $\hat{x} = i\hbar \partial/\partial p$ and $\hat{x}\hat{p}^{-1} - \hat{p}^{-1}\hat{x} = -i\hbar\hat{p}^{-2}$, one can prove that

$$x\hat{T}_{non} - \hat{T}_{non}x = -i\hbar(\hat{H}_{non}^{-1}x + x\hat{H}_{non}^{-1})/4. \tag{20}$$

where $\hat{H}_{non} = \hat{p}^2/2m$. Eq. (20) implies that there is an uncertainty relation between the time-of-arrival and position-of-arrival.

## Acknowledgments

The first author (Z. Y. Wang) would like to thank professor Haitang Yang for his useful discussions. This work was supported by the National Natural Science Foundation of China (Grant No. 60671030).

**Appendix A: Dervation of Eq. (11)**

Using Eq. (7) one has

$$c^2 t'^2 - x'^2 = [\frac{t\hat{H}}{mc} - \frac{(\hat{p}x + x\hat{p})}{2mc}]^2 - [\frac{(\hat{H}x + x\hat{H})}{2mc^2} - \frac{t\hat{p}}{m}]^2, \tag{a1}$$

Using $\hat{H}t = t\hat{H}$, $t\hat{p} = \hat{p}t$, $xt = tx$ and $\hat{H}\hat{p} = \hat{p}\hat{H}$ one has

$$\frac{1}{2m^2c^2}[-t\hat{H}(\hat{p}x + x\hat{p}) - (\hat{p}x + x\hat{p})t\hat{H} + (\hat{H}x + x\hat{H})t\hat{p} + t\hat{p}(\hat{H}x + x\hat{H})] = 0, \tag{a2}$$

then

$$c^2 t'^2 - x'^2 = \frac{t^2 \hat{H}^2}{m^2 c^2} + \frac{(\hat{p}x + x\hat{p})(\hat{p}x + x\hat{p})}{4m^2 c^2} - \frac{(\hat{H}x + x\hat{H})(\hat{H}x + x\hat{H})}{4m^2 c^4} - \frac{t^2 \hat{p}^2}{m^2}. \tag{a3}$$

Using $\hat{p}x = x\hat{p} - i\hbar$, $\hat{H}x = x\hat{H} - i\hbar\hat{H}^{-1}\hat{p}c^2$, and $\hat{H}^2 = \hat{p}^2 c^2 + m^2 c^4$, one has

$$c^2 t'^2 - x'^2 = c^2 t^2 + \frac{(2x\hat{p} - i\hbar)(2x\hat{p} - i\hbar)}{4m^2 c^2} - \frac{(2x\hat{H} - i\hbar\hat{H}^{-1}\hat{p}c^2)(2x\hat{H} - i\hbar\hat{H}^{-1}\hat{p}c^2)}{4m^2 c^4}. \tag{a4}$$

Because

$$(2x\hat{p} - i\hbar)(2x\hat{p} - i\hbar) = 4x\hat{p}x\hat{p} - 4i\hbar x\hat{p} - \hbar^2$$
$$= 4x(x\hat{p} - i\hbar)\hat{p} - 4i\hbar x\hat{p} - \hbar^2 = 4x^2\hat{p}^2 - 8i\hbar x\hat{p} - \hbar^2$$ , (a5)

$$(2x\hat{H} - i\hbar\hat{H}^{-1}\hat{p}c^2)(2x\hat{H} - i\hbar\hat{H}^{-1}\hat{p}c^2)$$
$$= 4x\hat{H}x\hat{H} - 2i\hbar x\hat{p}c^2 - 2i\hbar\hat{p}c^2\hat{H}^{-1}x\hat{H} - \hbar^2\hat{H}^{-2}\hat{p}^2c^4$$
$$= 4x(x\hat{H} - i\hbar\hat{H}^{-1}\hat{p}c^2)\hat{H} - 2i\hbar x\hat{p}c^2 - 2i\hbar\hat{p}c^2\hat{H}^{-1}(\hat{H}x + i\hbar\hat{H}^{-1}\hat{p}c^2) - \hbar^2\hat{H}^{-2}\hat{p}^2c^4$$
$$= 4x^2\hat{H}^2 - 4i\hbar x\hat{p}c^2 - 2i\hbar x\hat{p}c^2 - 2i\hbar\hat{p}c^2 x + 2\hbar^2\hat{H}^{-2}\hat{p}^2c^4 - \hbar^2\hat{H}^{-2}\hat{p}^2c^4$$ , (a6)
$$= 4x^2\hat{H}^2 - 6i\hbar x\hat{p}c^2 - 2i\hbar c^2(x\hat{p} - i\hbar) + \hbar^2\hat{H}^{-2}\hat{p}^2c^4$$
$$= 4x^2\hat{H}^2 - 8i\hbar x\hat{p}c^2 - 2\hbar^2c^2 + \hbar^2\hat{H}^{-2}\hat{p}^2c^4$$

one has

$$\frac{(2x\hat{p} - i\hbar)(2x\hat{p} - i\hbar)}{4m^2c^2} - \frac{(2x\hat{H} - i\hbar\hat{H}^{-1}\hat{p}c^2)(2x\hat{H} - i\hbar\hat{H}^{-1}\hat{p}c^2)}{4m^2c^4}$$
$$= \frac{1}{4m^2c^4}[(4x^2\hat{p}^2c^2 - 8i\hbar x\hat{p}c^2 - \hbar^2c^2) - (4x^2\hat{H}^2 - 8i\hbar x\hat{p}c^2 - 2\hbar^2c^2 + \hbar^2\hat{H}^{-2}\hat{p}^2c^4)]$$
$$= \frac{1}{4m^2c^4}[-4x^2m^2c^4 + \hbar^2c^2 - \hbar^2\hat{H}^{-2}\hat{p}^2c^4]$$ , (a7)
$$= -x^2 + \frac{1}{4m^2c^4}[\hbar^2c^2\hat{H}^{-2}(\hat{H}^2 - \hat{p}^2c^2)]$$
$$= -x^2 + \hbar^2c^2\hat{H}^{-2}/4$$

then one has

$$c^2t'^2 - x'^2 = c^2t^2 - x^2 + \hbar^2c^2\hat{H}^{-2}/4 ,$$ (a8)

which is exactly Eq. (11).